\documentclass{article}
\usepackage{amsmath}

\input{tcilatex}

\begin{document}

\begin{center}
{\LARGE Signal delay analysis for binary pulsars}

Trevor W. Marshall

CCAB, Cardiff University, 2 North RD., Cardiff CF10 1DY,UK
\end{center}

\textbf{Abstract\quad }This note gives a correction to the standard analysis
of the delay pattern in the radio signals from a pulsar in a binary system;
the same coordinate frame should be used for the transmission of the signal
as for the motion of the pulsar in the field of its companion.

\begin{center}
\ldots \ldots \ldots \ldots \ldots
\end{center}

\QTP{Body Math}
\bigskip The binary pulsar has been dubbed a "unique gravitational
laboratory"\cite{kramer}; the title is apt, because we are on the threshold
of observing the spin-orbit and tidal aspects of the system. But a necessary
prerequisite is that we have a correct description in the point-particle
approximation. The variable delay of a radio signal from a pulsar in a
binary system as it traverses its orbit has been analysed by Damour and
Taylor\cite{DT}, and their results continue to be used in observational
analysis up to the present time. The purpose of this note is to point out
that their article contains an error, due to different choices in the
coordinate frames used in the analyses of the Roemer and Shapiro delays
respectively.

The first analysis of the Shapiro delay was by Blandford and Teukolsky\cite%
{BT} (BT), but they did not propose its use in observational analysis,
because it is of the same order of magnitude as the correction, arising from
first post Newtonian (1pN) modification of the Kepler orbit, to the Roemer
delay. We begin by considering the case that the pulsar's mass is negligible
compared with that of its companion, so that the system in question is
planetary. A Kepler orbit is specified in the observer's coordinate system
by a set of five parameters, namely the longitude of the line of nodes $%
\Omega $, the inclination $i$ to the plane of the sky, the angle from the
line of nodes to the perihelion $\varpi $, the semimajor axis $a$, and the\
eccentricity $e.$ The position of the pulsar on its orbit is specified by
the polar angle $\phi $, measured from the line of nodes. Based on an
isotropic metric%
\begin{equation}
ds^{2}=\left( 1-\frac{2m}{r}\right) dt^{2}-\left( 1+\frac{2m}{r}\right) %
\left[ dr^{2}+r^{2}\left( d\theta ^{2}+\sin ^{2}\phi \right) \right] \quad ,
\end{equation}%
where terms of order $m^{2}$ have been discarded, the Roemer delay from a
point on the orbit is%
\begin{equation}
\Delta _{\text{R}}=r\sin i\sin \phi \quad ,  \label{Roemer}
\end{equation}%
and the Shapiro delay, which takes account of the varying refractive index
of space due to the companion's gravitational field and the consequent
bending of the signal's trajectory, was obtained by BT as%
\begin{equation}
\Delta _{\text{S}}=-2m\left[ \ln \frac{r}{a}+\ln \left( 1-\sin i\sin \phi
\right) \right] \quad .  \label{shapiro}
\end{equation}%
It should be noted that BT also calculated the Einstein delay time $\Delta _{%
\text{E}}$, which is the varying gravitational red shift, and which is
intermediate in order of magnitude; $\Delta _{\text{S}}/\Delta _{\text{R}}$
is of order $v^{2}/c^{2}$ while $\Delta _{\text{E}}/\Delta _{\text{R}}$ is
of order $v/c$. The total delay is the sum $\Delta _{\text{R}}+\Delta _{%
\text{E}}+\Delta _{\text{S}},$ and the maximum value of $v/c$ for pulsars
observed so far is about 2$\times 10^{-3}$.

The 1pN correction to $\Delta _{\text{R}}$ was given by Epstein\cite{Epstein}%
, the quantity $r$ in (\ref{Roemer}) being given, in the planetary limit, as%
\begin{equation}
r=a\left[ 1-\left( e+\Delta e\right) \cos \chi +f\cos 2\chi \right] \quad ,
\label{fourier}
\end{equation}%
where%
\begin{equation}
\tan \frac{\chi }{2}=\sqrt{\frac{1-e}{1+e}}\tan \frac{\lambda \left( \phi
-\varpi \right) }{2},\quad \lambda =1-\frac{3m}{a\left( 1-e^{2}\right) }%
\quad ,
\end{equation}%
and%
\begin{equation}
\Delta e=\frac{me}{4a\left( 1-e^{2}\right) ^{2}}\left( 13-2e^{2}\right)
\left( 2-e^{2}\right) ,\quad f=\frac{me^{2}}{4a\left( 1-e^{2}\right) ^{2}}%
\left( 13-2e^{2}\right) \quad .
\end{equation}%
While the parameter $\Delta e$ is just a small change in the ellipticity of
the orbit, the other parameter $f$ gives a distortion of the orbit. The
consequent corrections to $\Delta _{\text{R}}$ are indeed, as anticipated by
BT, comparable in magnitude with $\Delta _{\text{S}}$, but because the
latter contains $\Delta _{\text{R}}/a$ an analysis of the delay pattern from
a double pulsar has enabled Kramer and Wex\cite{kramer} to measure $a$ and $i
$ separately, whereas previous analysis based on expressions for $\Delta _{%
\text{R}}$ and $\Delta _{\text{E}}$ allowed only measurement of $a\sin i$.

However, if we use the same isotropic metric for the planetary motion as for
the signal transmission, it is easy to deduce that to order $m$ the only
relativistic correction is in the constant precession rate%
\begin{equation}
\dot{\varpi}=\frac{3}{1-e^{2}}\sqrt{\frac{m^{3}}{a^{5}}}\quad ,
\label{precess}
\end{equation}%
that is there is no distortion of the elliptic orbit. If, for example, we
use the harmonic coordinates advocated by Fock\cite{fock}, as a precise
version of such a metric, that is%
\begin{equation}
ds^{2}=\frac{r-m}{r+m}dt^{2}-\frac{r+m}{r-m}dr^{2}-\left( r+m\right)
^{2}\left( d\theta ^{2}+\sin ^{2}\theta d\phi ^{2}\right) \quad ,
\end{equation}%
then the relativistic orbit in the plane $\theta =\pi /2$ may be obtained
from the well known equation for the Schwarzschild radial coordinate $R$,
namely\cite{eddington}%
\begin{equation}
\left( \frac{dR}{d\phi }\right) ^{2}=\frac{E^{2}-1}{J^{2}}R^{4}+\frac{2m}{%
J^{2}}R^{3}-R^{2}+2mR\quad ,
\end{equation}%
where $E$ and $J$ are the energy and angular momentum integrals, simply by
putting $R=r+m$, giving, to 1pN order,%
\begin{equation}
\left( \frac{dr}{d\phi }\right) ^{2}=\frac{E^{2}-1}{J^{2}}r^{4}+\frac{2m}{%
J^{2}}\left( 2E^{2}-1\right) r^{3}-\lambda ^{2}r^{2},\quad \lambda =1-\frac{%
3m^{2}}{J^{2}}\quad .
\end{equation}%
Then, defining new constants $a$ and $e$ by%
\begin{equation}
E=1-\frac{m}{2a}+\frac{7m^{2}}{8a^{2}},\quad J=\frac{\sqrt{2m\left(
1-e^{2}\right) }}{\lambda }\left( 1-\frac{m}{a}\right) \quad ,
\end{equation}%
this factorizes as%
\begin{equation}
\left( \frac{dr}{d\phi }\right) ^{2}=\frac{\lambda ^{2}r^{2}}{a^{2}\left(
1-e^{2}\right) }\left( r-a+ae\right) \left( a+ae-r\right) \quad ,
\end{equation}%
so, in harmonic coordinates, the orbit is%
\begin{equation}
r=\frac{a\left( 1-e^{2}\right) }{1+\cos \psi },\quad \psi =\lambda \left(
\phi -\varpi \right) \quad ,
\end{equation}%
which is the equation of an ellipse precessing, without distortion, at the
rate (\ref{precess}). Note that $\phi $, as always, is measured from the
line of nodes, while $\psi $ is measured from an initial periastron;
successive periastrons occur at intervals of 2$\pi $.When we take account of
the slightly different expressions for $a$ and $e$ in the Schwarzschild
description, it gives an orbit similar to the Epstein orbit. But putting $R$
instead of $r$ into the expression for the Roemer delay gives an answer
which contains the additional term $m\sin i\sin \phi $. It may be verified
that the Shapiro delay, calculated in the Schwarzschild instead of the
isotropic coordinates of BT contains another additional term which exactly
cancels this one. The expression given by Epstein\cite{Epstein} and by
Damour and Taylor\cite{DT}, which combines a calculation of the Roemer delay
in the coordinates of Einstein, Infeld and Hoffman with one of the Shapiro
delay in the isotropic coordinates, is incorrect.

In order to obtain the correct delay formula we now extend the above
analysis to the case of two masses of comparable magnitude. For this system
Fock\cite{fock} derives the orbit from a lagrangian (see his eqn. (81.01)
simplified by using (81.18)) 
\begin{equation}
L=\frac{1}{2}v^{2}+\frac{m}{r}+\frac{3v^{4}}{8}\left( 1-\frac{3m^{\ast }}{m}%
\right) +\frac{v^{2}\left( 3m+m^{\ast }\right) }{2r}-\frac{m^{2}}{2r^{2}}-%
\frac{m^{\ast }}{2r}\dot{r}^{2}
\end{equation}%
to give the energy and angular momentum integrals%
\begin{eqnarray}
E_{1} &=&\frac{1}{2}v^{2}-\frac{m}{r}+\frac{3}{2}E_{1}^{2}\left( 1-\frac{%
3m^{\ast }}{m}\right) +\frac{E_{1}\left( 6m-7m^{\ast }\right) }{2r}  \notag
\\
&&+\frac{m\left( 10m-5m^{\ast }\right) }{2r^{2}}-\frac{1}{2}m^{\ast }r\dot{%
\phi}^{2}
\end{eqnarray}%
and%
\begin{equation}
J=r^{2}\dot{\phi}\left[ 1+E_{1}\left( 1-\frac{3m^{\ast }}{m}\right) +\frac{%
4m-2m^{\ast }}{r}\right] \quad .  \label{Jint}
\end{equation}%
These combine to give the orbit equation%
\begin{align}
\left( \frac{dr}{d\phi }\right) ^{2}& =\frac{r^{4}}{J^{2}}\left[
2E_{1}+E_{1}^{2}\left( 1-\frac{3m^{\ast }}{m}\right) \right] +\frac{2mr^{3}}{%
J^{2}}\left( 1+4E_{1}-\frac{3m^{\ast }E_{1}}{m}\right)  \notag \\
& -r^{2}\left( 1-\frac{6m^{2}}{J^{2}}+\frac{3mm^{\ast }}{J^{2}}\right)
+m^{\ast }r\quad ,
\end{align}%
where%
\begin{equation}
m=m_{1}+m_{2},\quad m^{\ast }=\frac{m_{1}m_{2}}{m_{1}+m_{2}}\quad .
\end{equation}%
Now we define%
\begin{equation}
\lambda =1-\frac{3m^{2}}{J^{2}},\quad r=R+\frac{1}{2}m^{\ast }\quad ,
\end{equation}%
giving%
\begin{equation}
\left( \frac{dR}{d\phi }\right) ^{2}=\frac{R^{4}}{J^{2}}\left[
2E_{1}+E_{1}^{2}\left( 1+\frac{3m^{\ast }}{m}\right) \right] +\frac{2mR^{3}}{%
J^{2}}\left( 1+4E_{1}-\frac{m^{\ast }E_{1}}{m}\right) -\lambda
^{2}R^{2}\quad ,
\end{equation}%
which, putting%
\begin{equation}
\lambda J=\sqrt{ma\left( 1-e^{2}\right) }\left[ 1-\frac{4m+m^{\ast }}{4a}%
\right] ,\quad E_{1}=-\frac{m}{2a}+\frac{7m^{2}+mm^{\ast }}{a^{2}}\quad ,
\end{equation}%
gives the same orbit equation as in the planetary case, but with $r$
replaced by $R$. Note that $a$ and $e$ reduce to their planetary values on
putting $m^{\ast }=0$, and $E=1+E_{1}$.Thus the general $r$ is simply 
\begin{equation}
r=\frac{a\left( 1-e^{2}\right) }{1+e\cos \psi }+\frac{1}{2}m^{\ast },\quad
\psi =\lambda \left( \phi -\varpi \right) \quad .
\end{equation}

The rate at which the orbit is described is obtained from the angular
momentum integral (\ref{Jint}), leading to%
\begin{eqnarray}
t &=&\int \frac{r^{2}}{\lambda J}\left( 1-\frac{m-3m^{\ast }}{2a}+\frac{%
4m-2m^{\ast }}{r}\right) d\psi  \notag \\
&=&\sqrt{\frac{a^{3}}{m\left( 1-e^{2}\right) }}\int \frac{r^{2}}{a^{2}}%
\left( 1+\frac{2m+5m^{\ast }}{4a}+\frac{4m-2m^{\ast }}{r}\right) d\psi \quad
.
\end{eqnarray}%
Then, substituting the orbit equation for $r$,%
\begin{equation}
t=\sqrt{\frac{a^{3}}{m}}\int \frac{\left( 1-e^{2}\right) ^{3/2}}{\left(
1+e\cos \psi \right) ^{2}}\left[ 1+\frac{2m+5m^{\ast }}{4a}+\frac{4m-m^{\ast
}}{a\left( 1-e^{2}\right) }\left( 1+e\cos \psi \right) \right] d\psi
\end{equation}%
giving the secular equation%
\begin{equation}
nt=\chi -e\left( 1-\frac{4m-m^{\ast }}{a}\right) \sin \chi \quad ,
\end{equation}%
where%
\begin{equation}
\chi =2\tan ^{-1}\left( \sqrt{\frac{1-e}{1+e}}\tan \frac{\psi }{2}\right)
\quad ,
\end{equation}%
and $n$ is the frequency%
\begin{equation}
n=\sqrt{\frac{m}{a^{3}}}\left( 1-\frac{18m+m^{\ast }}{4a}\right) \quad .
\end{equation}%
This equation may be inverted to give $\psi ,$ and hence $\phi $ and $r$, as
a function of $t$. Note that both the orbit and its rate of description, in
the harmonic coordinates, are remarkably similar to their Newtonian forms$.$
We may finally express the Roemer delay%
\begin{equation}
\Delta _{R}=r\sin \phi \sin i
\end{equation}%
as a function of $t$, and this is the quantity which must be added to the BT
values of $\Delta _{\text{E}}$ and $\Delta _{\text{S}}$ to give the total
delay.


\begin{thebibliography}{9}
\bibitem{kramer} M. Kramer and N. Wex, \emph{Class. Quantum Gravit. }\textbf{%
26,} 073001 (2009)

\bibitem{DT} T. Damour and J. H. Taylor, \emph{Phys. Rev. D }\textbf{45, }%
1840 (1992)

\bibitem{BT} R. Blandford and S. A. Teukolsky, \emph{Ap. J. }\textbf{205},
580 (1975)

\bibitem{Epstein} R. Epstein, \emph{Ap. J. }\textbf{216,} 92-100 (1977)

\bibitem{eddington} A. S. Eddington, \emph{The Mathematical Theory of
Relativity, }para 40\emph{, }University Press, Cambridge (1924)

\bibitem{fock} V. Fock, \emph{The Theory of Space, Time and Gravitation, 2nd
ed.,} Pergamon, Oxford (1966)
\end{thebibliography}
\end{document}